\documentclass[aps,prl,twocolumn,superscriptaddress,10pt]{revtex4-1}

\usepackage{graphicx}
\usepackage{amsmath}
\usepackage{amssymb}
\usepackage{amsfonts}
\usepackage{bm}
\usepackage{psfrag}
\usepackage{pstricks}
\usepackage{color}
\usepackage{soul}
\usepackage{tabularx}
\usepackage{nicefrac}
\usepackage[version=4]{mhchem}

\pdfoutput=1

\usepackage{hyperref}
\hypersetup{colorlinks=true, linkcolor=blue, citecolor=blue, urlcolor=blue}

\usepackage[normalem]{ulem} 
\usepackage{cancel}
\let\isout\sout 
\renewcommand{\sout}[1]{\ifmmode\text{\isout{\ensuremath{#1}}}\else\isout{#1}\fi}

\begin{document}

\widetext
\title{Strain-engineered divergent electrostriction in \ce{KTaO3}}


\author{Daniel S.~P.~Tanner}
\email{danielsptanner@gmail.com} 
\affiliation{Universit\'e de Li\`ege, Q-MAT, CESAM, Institut de Physique}
\affiliation{Universit\'e Paris-Saclay, CentraleSup\'elec, CNRS, Laboratoire SPMS, 91190 Gif-sur-Yvette, France}
\author{Pierre-Eymeric Janolin}
\affiliation{Universit\'e Paris-Saclay, CentraleSup\'elec, CNRS, Laboratoire SPMS, 91190 Gif-sur-Yvette, France}
\author{Eric Bousquet}
\affiliation{Universit\'e de Li\`ege, Q-MAT, CESAM, Institut de Physique}
\email{email} 

\date{\today}


\begin{abstract}

We investigate the electrostrictive response across a ferroelectric phase
transition from first-principles calculations and 
show that $M$, the field-induced electrostrictive tensor, controlling the amplitude of the electric-field induced strain, can be made arbitrarily large through strain engineering.
We take as a case study the epitaxial strain-induced transition from para- to ferro-oelectricity of \ce{KTaO3}.
We show that the magnitude of the field-induced electrostriction diverges with the permitivity at the transition, hence exhibiting giant responses through a calculation of both the $M$ and $Q$ electrostrictive tensors. 
We explain the origin of this giant electrostrictive response in \ce{KTaO3} using a microscopic decomposition of the electrostriction coefficients, and use this understanding to propose design rules for the development of future giant electrostrictors for electromechanical applications. Finally, we introduce a further means to calculate electrostriction, specific to ferroelectrics, and not yet utilised in the literature.
\end{abstract}

\maketitle

Electrostriction is an electromechanical coupling present in all dielectrics which describes the quadratic induction of a strain or stress, due to an electric or polarisation field.
Despite its ubiquity, it is often dwarfed in amplitude  by its linear counterpart, piezoelectricity, which constitutes the primary response in most electromechanical systems. 
However, renewed interest in electrostriction has grown due to the recent discovery of materials which exhibit "giant" electrostrictive responses,\cite{Korobko2012,Li2018,Yuan2018} capable of competing with piezoelectrics in terms of responses.
These electrostrictors exhibit a host of attractive advantages over piezoelectrics such as: 
(i) minimal hysteresis in the strain-field response, 
(ii) increased stability of induced deformations,
(iii) smaller response time,
(iv) enhanced temperature stability, and
(v) materials which are lead-free and environmentally friendly.\cite{Cieminski1991,KaIm17,LiJi14}

In ferroelectric materials, there is a consensus that the large electromechanical response of ferroelectric materials is mostly due to  their piezoelectric rather than electrostrictive response. 
Additionally, it is often accepted from experimental studies~\cite{KaIm17,UwSa75,LiJi14} and phenomenological treatment~\cite{levanyuk1998} that the electrostriction remains relatively small and unchanged across ferroelectric phase transitions. 
However, these studies report the electrostrictive $Q_{ijkl}$ coefficient (relating strain to the polarisation squared) while a few reports \cite{IsSm89, ujma1985} 
found that the electrostrictive $M_{ijkl}$ tensor (relating strain to the electric field squared) increases rapidly with temperature together with the permittivity near 
phase transitions. However, it is believed that the $M_{ijkl}$ response is nevertheless negligibly small with respect to the piezoelectric response, and therefore useless in terms of technological applications \cite{shuvalov1988}.
Hence, the question remains 
of whether inducing phase transitions where the permittivity diverges enables to increase the electrostrictive response significantly enough to make electrostriction interesting for electromechanical applications. 


In this Letter, through a thorough examination of electrostriction at the strain-induced ferroelectric phase transition (FEPT) in \ce{KTaO3} from first-principles calculations, we demonstrate a giant electrostrictive response can indeed be obtained in ferroelectric materials at the phase transition, confirming that, though the components of the polarisation tensor $Q$ do not change much, the out-of-plane electric-field tensor components diverge and reaches values that make electrostriction interesting for electromechanical applications.
In addition, through a microscopic decomposition of the electrostriction tensors, we show that it is the soft transverse optical mode that is responsible for the large 
electrostriction. 

To proceed, we note that while one may consider electrostriction as the ``strain/stress induced quadratically by an electric/polarisation field'', one may also view it as the theromodynamically equivalent ``linear change in susceptibility/inverse susceptibility induced by a stress or a strain''~\cite{Devonshire,TaBo21} shown below:
\begin{equation} \label{eq:Electro}
 \begin{aligned}
  &\frac{1}{\epsilon_{0}}\frac{\partial\eta_{ij}}{\partial X_{kl}} = -2Q_{ijkl}   &\frac{1}{\epsilon_{0}}\frac{\partial\eta_{ij}}{\epsilon_{0}\partial x_{kl}} = 2q_{ijkl} \\
  &\epsilon_{0}\frac{\partial\chi_{ij}}{\partial X_{kl}} = 2M_{ijkl} 
  &\epsilon_{0}\frac{\partial\chi_{ij}}{\partial x_{kl}} = -2 m_{ijkl}.
  \end{aligned}
\end{equation}
See the supplemental material for details of the theromodynamical derivation of these equations.~\footnote{See Supplemental Material at [URL will be inserted by
  publisher] for thermodynamical equivalence with expressions of direct electrostriction.} In our previous work~\cite{TaBo21}, we outlined the advantages of this method of calculation, amongst which are greater computational efficiency and robustness.

Viewed from the perspective of Eqs.\eqref{eq:Electro}, we see that at a strain- or stress-induced phase transition, involving the divergence of the susceptibility, \textit{some} electrostrictive tensor components \textit{must} become large.~\cite{jona1962} 
To compute Eqs.~\eqref{eq:Electro} we have used the DFT package ABINIT (version 8.6.1) \cite{gonze2016}, with k-point grid densities of 8$\times$8$\times$8 
and a plane wave cutoff energy of 75 Ha, to ensure convergence of electrostrictive coefficients of about 1\%.
The PseudoDojo~\cite{Setten2018} normconserving pseudopotentials were used, and the exchange-correlation functional was treated using the generalised gradient approximation of Purdue, Burke, and Ernzerhof, modified for solids (PBEsol).\cite{PeRu08}
We note that obtaining the electrostriction tensors in this manner using eq.\ref{eq:Electro} and finite differences, will yield an "improper" electrostriction tensor containing  contributions described by Nelson and Lax~\cite{NeLa73}; the "proper" tensor may be obtained from this by the addition of permitivity components to certain electrostriction components.\cite{NeLa73,TaBo21}

As a test case, we will analyse the electrostrictive response at the strain-induced FEPT present in \ce{KTaO3}.
\ce{KTaO3} is a highly suitable material for such a study as it is an incipient  ferroelectric which undergoes a ferroelectric phase transition due to relatively small axial stress.
A FEPT is reported at 0.6\,GPa by Uwe~\cite{UwSa75} and for a  strain of ~1\% by Tyunina~\cite{TyNa10} at low temperature (4.2\,K), which can be studied experimentally on single crystals or epitaxial thin films. 
In addition, \ce{KTaO3} has a wide range of applications~\cite{TyNa10,Taganstsev2000,UeNa11,Liu716,GeMa21}, including the recent discoveries of superconductivity~\cite{UeNa11,Liu716} or dynamic multiferroism~\cite{GeMa21}, making it a highly technologically relevant material.

We present in Table~\ref{tab:CompAndExp} our calculated ground state properties of \ce{KTaO3} along with previous PBEsol calculations from the literature, as well as experimental measurements. 
We obtain a bandgap in good agreement with previous PBEsol theory, but it is underestimated with respect to the experimental values as expected from the PBEsol functional, though we can reproduce the indirect gap from $R$ point to $\Gamma$-point. 
For the lattice and elastic constants good agreement is obtained with both experiment and previous theory. 
While no previous experimental or theoretical works have determined the $M_{ijkl}$ electrostrictive coefficients, for the $Q_{ijkl}$ coefficients, we obtain reasonable agreement with previous experiment~\cite{UwSa75,LiJi14}, with the disagreement between our computed values and the experimental ones being equivalent to the variances amongst the experimental values.



\begin{table}[h] 
\centering  
\begin{tabularx}{\columnwidth}{>{\setlength\hsize{1\hsize}\centering}X c >{\setlength\hsize{1\hsize}\centering}X c }
\hline\hline
 Coefficient             
 & 
 Present Calc
 & 
 Lit Calc      
 & 
 Exp       
 \\ 
 \hline
    $E_{g}^{R-\Gamma}$~(eV)      & 2.16   & 2.14\textsuperscript{c}   & 4.35\textsuperscript{f}     \\ 
    $E_{g}^{\Gamma-\Gamma}$~(eV) & 2.77   & 2.70\textsuperscript{d}   & 3.64\textsuperscript{f}     \\ \hline
    $a_{0}$ (\AA)                & 3.9817 & 3.989\textsuperscript{c}  & 3.9885\textsuperscript{e}    \\ 
    $C_{11}$~(GPa)               & 453.2  & 474.7 \textsuperscript{c}  & 431 \textsuperscript{g}                 \\ 
    $C_{12}$~(GPa)               & 76.8   & 62.752\textsuperscript{c}  & 103 \textsuperscript{g}                 \\ 
    $C_{44}$~(GPa)               & 99.7   & 197.10 \textsuperscript{c} & 109 \textsuperscript{g}              \\ \hline
    $\omega_{TO1}$    &  63       &   104.01\textsuperscript{h}   & 85 \textsuperscript{i} 81\textsuperscript{j} \\
    $\omega_{LO1}$    & 171       &   173.08\textsuperscript{h}   & 185 \textsuperscript{j}    \\
    $\omega_{TO2}$    & 180       &   179.57\textsuperscript{h}   & 198 \textsuperscript{i} 199\textsuperscript{j} \\
    $\omega_{LO2}$    & 398       &   392.64\textsuperscript{h}*  &     \\
    $\omega_{TO3}=\omega_{LO3}$    & 256       &   237.7\textsuperscript{h}    & 279\textsuperscript{j}    \\     
    $\omega_{TO4}$    & 529       &   561.33\textsuperscript{h}     & 556 \textsuperscript{i} 546\textsuperscript{j} \\
    $\omega_{LO4}$    & 791       &   821.49\textsuperscript{h}     & 826\textsuperscript{j}    \\
    $\epsilon$      & 433       &                                 &                       
    \\
    \hline
    $Q_{11}$~(m$^{4}$C$^{-2}$)   & 0.1256  & -       & 0.087\textsuperscript{a}                 \\ 
    $Q_{12}$~(m$^{4}$C$^{-2}$)   & -0.0313 & -       & -0.023\textsuperscript{a}                 \\ 
    $Q_{h}$~(m$^{4}$C$^{-2}$)    & 0.063   & -       & 0.041\textsuperscript{a},0.052\textsuperscript{b}             \\ \hline
    $M_{11}$~(m$^{2}$V$^{-2}$)   & 1.98x10$^{-18}$  & -  & -                 \\ 
    $M_{12}$~(m$^{2}$V$^{-2}$)   & -4.70x10$^{-19}$ & -  & -                 \\ 
    $M_{h}$~(m$^{2}$V$^{-2}$)    & 1.05x10$^{-18}$  & -  & -              \\ \hline    
\hline
\end{tabularx}
\caption{Bulk paraelectric ground state  \ce{KTaO3} electronic, structural, elastic, and electrostrictive properties obtained via 
calculation and measurement. a=Ref[\onlinecite{UwSa75}]; b=Ref[\onlinecite{LiJi14}];   c=Ref[\onlinecite{BoHi2013}]; d=Ref[\onlinecite{BeBe12}; e=Ref[\onlinecite{Vousden1951}]; f=Ref[\onlinecite{JePa06}]; g=Ref[\onlinecite{WeberHB}]; h=Ref[\onlinecite{Suleyman2010}]; i=Ref[\onlinecite{FlWo68}] (soft mode at 300K, others at 10K); j=Ref[\onlinecite{VoUw84}] (RT).  
*In \textsuperscript{h} the LO2 mode is mislabled as the TO3 mode due to the fact that LO2 is higher in frequency than the non polar TO3 and LO3 modes.
} 
\label{tab:CompAndExp}
\end{table} 

To investigate electrostriction through the FEPT, we have calculated the properties of \ce{KTaO3} over a range of biaxial compressive and tensile strains from -0.7\% to 0.7\%, as may be achieved through epitaxial growth on various substrates. This was done by imposing the strain components $x_{11}=x_{22}=(a_s-a_0)/a_0$ where $a_0$ is the relaxed lattice constant of KTaO$_3$ and $a_s$ is the lattice constant of the substrate, and by relaxing the $z$ direction of strain. 

\begin{figure}[ht]
   \centering
   \includegraphics[width=\columnwidth]{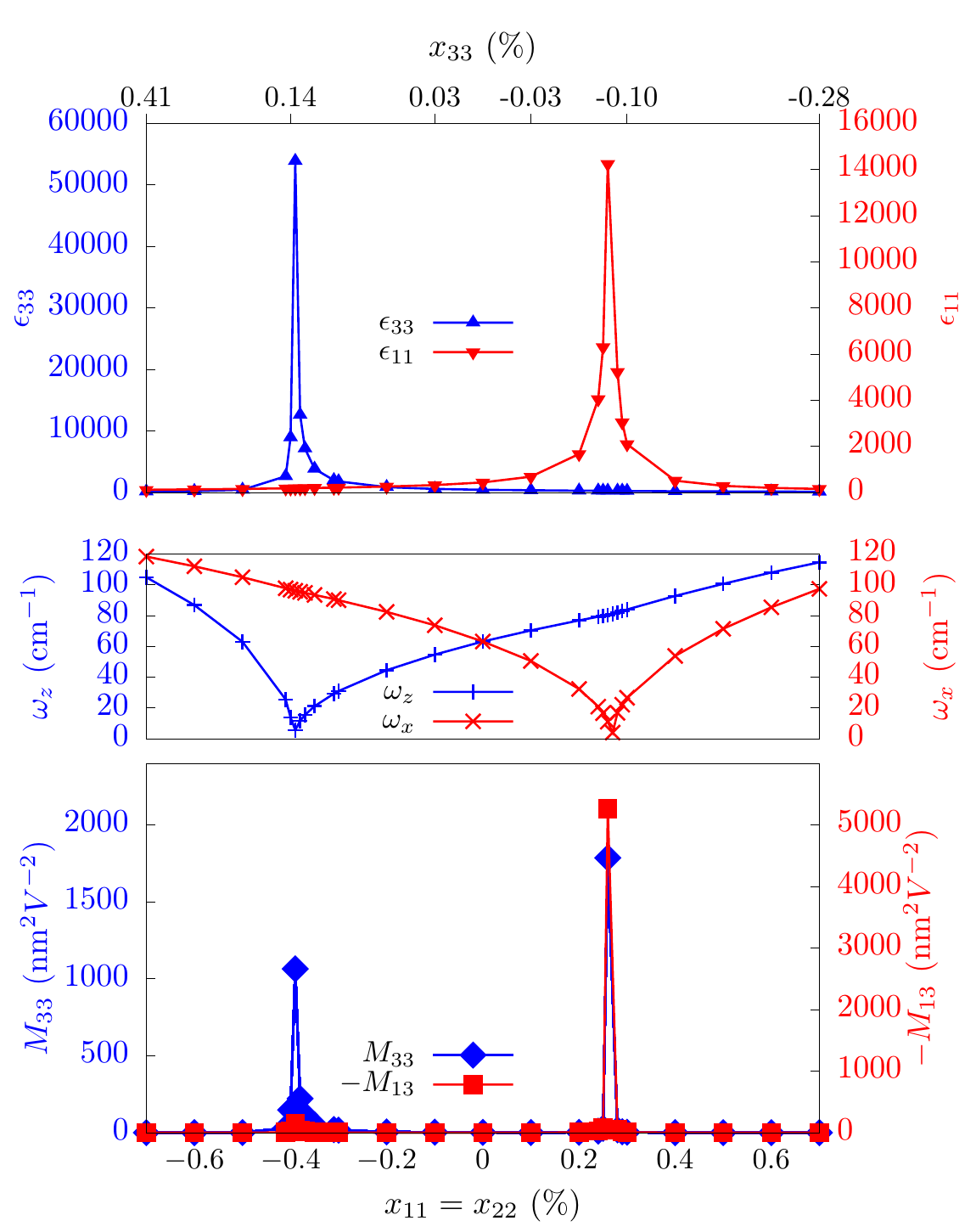}
   \caption{Variation with epitaxial strain ($x_{11}=x_{22}; X_{33}=0$) of: permitivity tensor components $\epsilon_{11}$ and $\epsilon_{33}$ (top); lowest x and z polar phonon modes, $\omega_{x}$ and $\omega_{z}$ (mid); and electrostriction tensor components $M_{33}$ and $M_{13}$ (bottom). Compressive epitaxial strain induces a tensile strain induced phase transition in the z-direction, and tensile epitaxial strain induces a phase transition in the x=y direction.
  }
   \label{fig:ElectroEpitax}
\end{figure}

In the top part of Fig.~\ref{fig:ElectroEpitax} we show the impact of these various epitaxial strains on the permitivity tensor. 
Consistent with previous works on \ce{KTaO3}, the epitaxial strain induces a FEPT~\cite{UwSa75,TyNa10}, and we see it is in fact the tensile strains which drive the polar displacement. 
On the left, the middle panel reveals that the tensile strain induced in the z-direction by the compressive epitaxial xy strain leads to the softening of the transverse optic (TO) mode polar in $z$, and a divergence of $\epsilon_{33}$, with the FEPT occurring at 0.4\% epitaxial xy strain; on the right, the epitaxial tensile strain leads to the simultaneous softening of the polar TO modes in $x$ and $y$, and a divergence of $\epsilon_{11}$ and $\epsilon_{22}$, with the transition occurring at around 0.26\% epitaxial strain. This earlier transition point is due to the fact that in this case the tensile strain is being directly imposed on the x and y coordinates, whereas with the compressive strain, the tensile strain in the z-direction occurs more slowly due to a Poisson effect. This occurrence of the FEPT at the rather small compressive and tensile strains of 0.4 and 0.26\% is consistent with the incipient nature of the FE in \ce{KTaO3}.

The bottom panel of Fig.~\ref{fig:ElectroEpitax} shows that the electrostrictive tensor components $M_{33}$ and $M_{13}$ also diverge at the FEPT, contrary to reports in the literature that electrostriction is constant at a phase transition~\cite{LiJi14}. To compute the electrostriction, at each fixed epitaxial strain, we apply a further strain in the z-direction, of magnitude of $-0.005$, to the already relaxed c lattice vector, and calculate again the permitivity using DFPT. The coefficient $m_{33}$ is obtained from the derivative of $\epsilon_{z}$ with respect to the applied $x_{33}$; and the rate of change of $\epsilon_{x}$ w.r.t. $x_{33}$ yields the coefficient $m_{13}$. The technologically relevant $M_{33}$ and $M_{13}$, are then derived by appropriate consideration of the epitaxial boundary conditions. With:  
\begin{equation} \label{eq:mtoM}
  \begin{aligned}
   &M_{33}=2s_{12}m_{13}+s_{33}m_{33} \\
   &M_{13}=(s_{11}+s_{12})m_{13}+s_{13}m_{33}, 
  \end{aligned}
\end{equation}
where $s_{ij}$ are elastic susceptibilities, for which we have used the equilibrium values,
noting that epitaxial strains of $<\pm1$ are unlikely to induce significant elastic non-linearities~\cite{TaCa19}.

Focusing first on the compressive epitaxial strain induced transition (at -0.4\%), we observe spikes in both $M_{33}$ and $M_{13}$; with $M_{33}$
reaching 1x10$^{-15}$\,m$^{4}$/C$^{2}$, a value rarely observed except for some polymers, and $M_{13}$ reaching the smaller, though still considerable value of -1.5x10$^{-16}$\,m$^{4}$/C$^{2}$. This means that in a compressively strained epitaxial layer, an electric field will produce a significant tensile strain in the z-direction, and also a compressive strain parallel to the plane of the layer. To put this in context, the peak $M_{33}$ is equivalent to an \textit{effective} piezoelectric coefficient of $d_{33}^{eff}$=~$60,000$~pm/V, which compares very favourably with $d_{33}$ of up to 600\,pm/V~\cite{AlCh01} values in classic piezoelectric material PZT. This large tensile z-strain could have applications in micro-fluidics, adjustable mirrors etc., where the lead-free nature of \ce{KTaO3} makes it particularly attractive when compared with other lead-containing materials.

Looking at the tensile-strain-induced transition (at 0.26\%), we find again a divergence of the coefficients, though this time both reaching much higher values, with $M_{13}$ reaching -5x10$^{-15}$\,m$^{4}$/C$^{2}$, and $M_{33}$ reaching 2x10$^{-15}$\,m$^{4}$/C$^{2}$. 
In this case an electric field in z will produce considerably more strain in-plane/xy than in the case of compressive epitaxial strain. 

Figure~\ref{fig:ElectroEpitax} thus demonstrates that electrostrition in lead-free \ce{KTaO3}, brought near the phase transition through epitaxial growth on an appropriate substrate, may be an ideal active material candidate for future electromechanical devices. In addition, we note that the values obtained of the order of 10$^{-15}$\,m$^{4}$/C$^{2}$, compare very favorably with known giant electrostrictors, such as  (Nb,Y)-doped \ce{Bi2O3} (M$\approx$10$^{-17}$\,m$^{4}$/C$^{2}$)~\cite{YaSm18}, Gd-doped ceria (M$\approx$10$^{-17}$ to 10$^{-16}$\,m$^{4}$/C$^{2}$)~\cite{YaYe18} and \ce{La2Mo3O9} (M=1.5x10$^{-18}$\,m$^{4}$/C$^{2}$)~ \cite{JiachPierre} without being restricted to low frequencies.

\begin{figure}[h]
   \centering
   \includegraphics[width=\columnwidth]{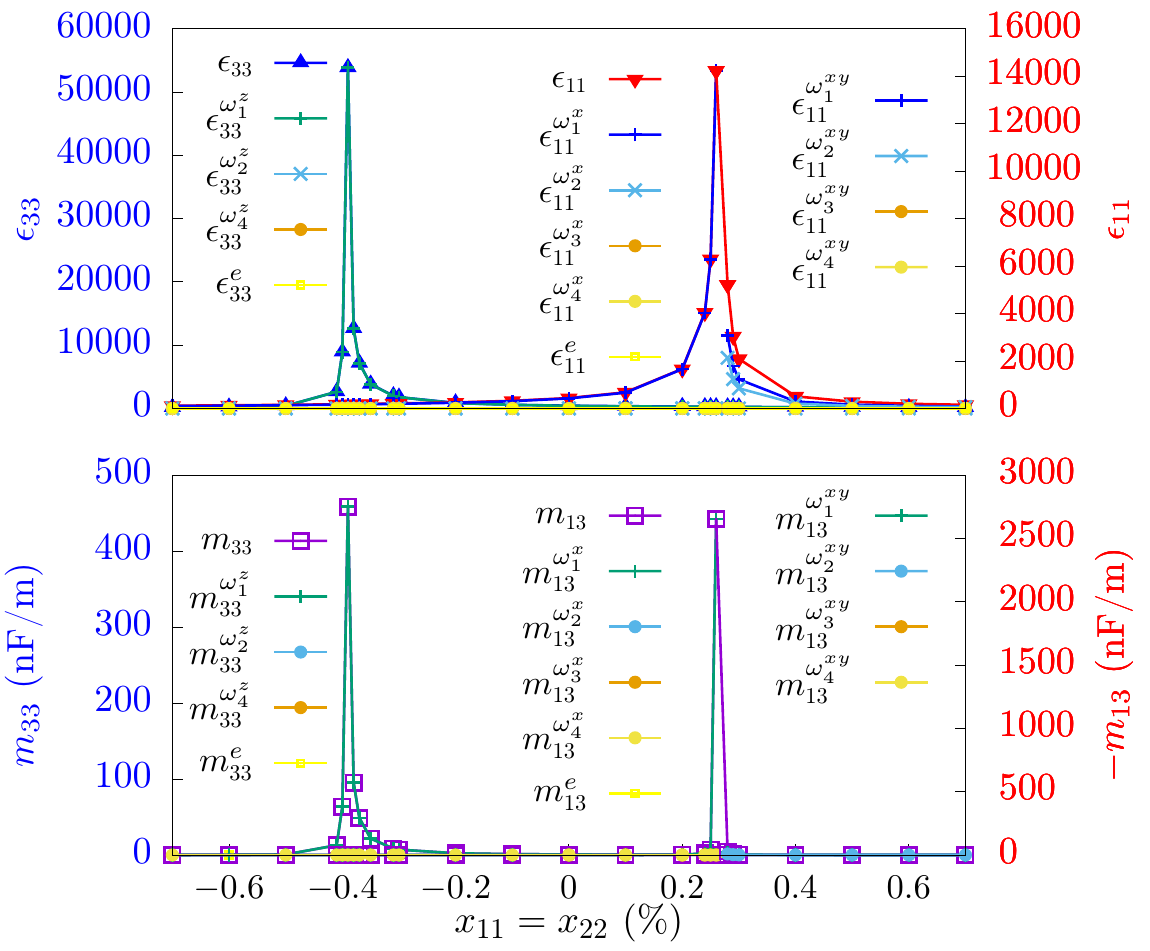}
   \caption{Decomposition of (top) permitivity tensor components $\epsilon_{33}$ and $\epsilon_{11}$, and (bottom) electrostrictive tensor components $m_{33}$ and $m_{13}$ into their mode by mode and electronic contributions. For both the permitivity and electrostriction, the primary contribution is from the softening TO mode.}
   \label{fig:Decomp}
\end{figure}

To better understand this giant electrostriction, we can profit from the particular manner in which the methods based on eqs.\ref{eq:Electro}, and introduced in Ref.\onlinecite{TaBo21}, determine the electrostrictive coefficients. Through the decomposition of the permitivity tensor into its mode by mode contributions, we are also able to decompose the electrostriction tensor. This decomposition is illustrated for the $\epsilon_{x}$,$\epsilon_{z}$ and the electrostriction tensor components $m_{33}$ and $m_{13}$ in Fig.~\ref{fig:Decomp}. The figure shows that all four TO modes contribute to $\epsilon_{11}$  and $m_{13}$, whilst only three contribute to $\epsilon_{33}$  and $m_{33}$. However, amongst all of these contributions, the only that matter near the transitions are those of the soft TO1 modes, with all others being negligible and indistinguishable from zero in the plot. We note also that in Fig.~\ref{fig:Decomp}, the behaviour of the $m_{ij}$ is much simpler than that of the corresponding $M_{ij}$ shown in the bottom panel of Fig.~\ref{fig:ElectroEpitax}, with $m_{33}$ and $m_{13}$ each exhibiting only one peak; however, we see that it is the mixing of these coefficients, according to Eq.\ref{eq:mtoM}, which leads to the double peaked $M_{ij}$. 

\begin{figure}[h]
   \centering
   \includegraphics[width=\columnwidth]{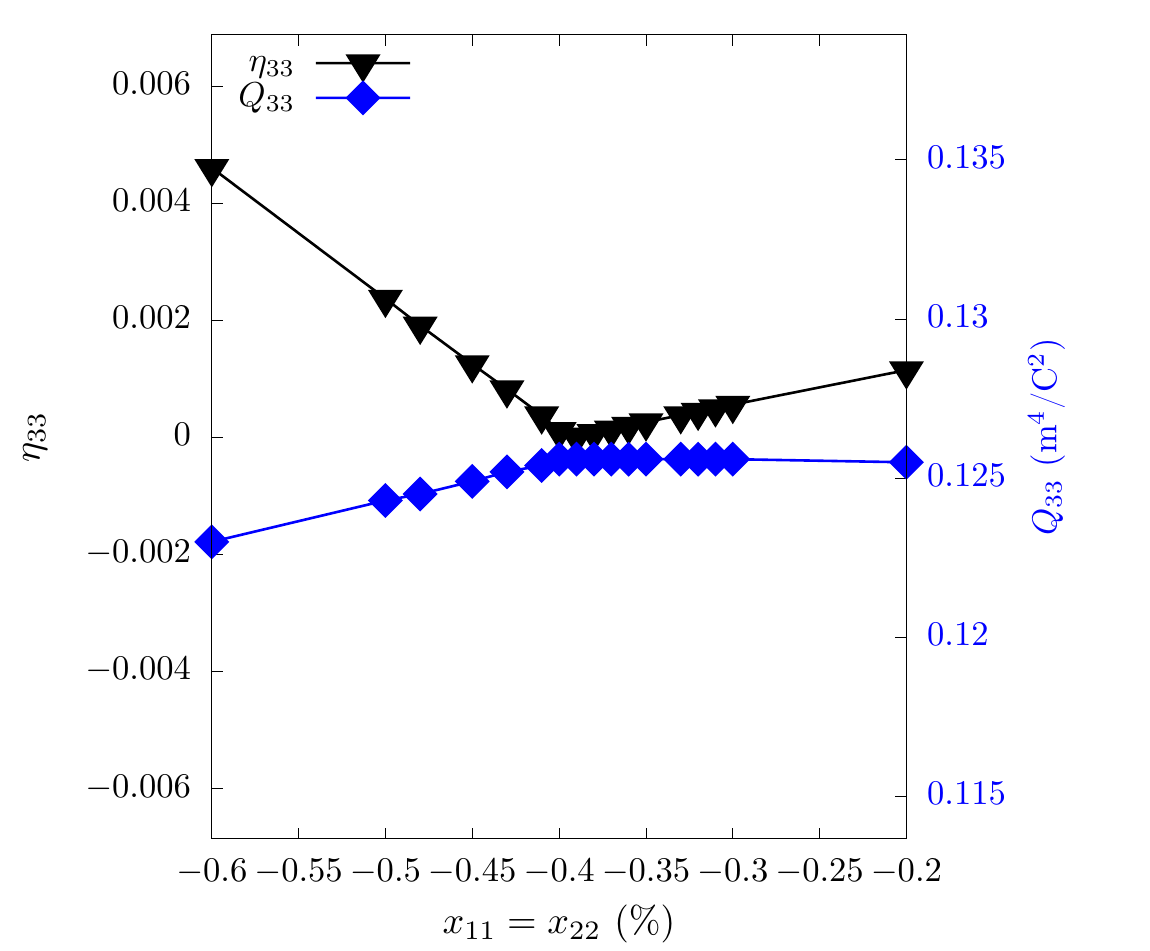}
   \caption{$\eta_{33}$ and $Q_{33}$ plotted against epitaxial strain. $Q_{33}$ is obtained by applying a further z-strain of -0.05\% at each epitaxial strain and computing the rate of change of $\epsilon_{33}$ with respect to stress. See text.}
   \label{fig:Q33Epitax}
\end{figure}

To clarify why some previous experimental observations reported that the $Q$ electrostrictive response is constant across FEPT,\cite{LiJi14} we have also examined the evolution of the $Q_{33}$ coefficient of our strained KTaO$_3$. Figure~\ref{fig:Q33Epitax} shows the development of $\eta_{33}$ and $Q_{33}$ at the compressive strain induced phase transition. We plot $\eta_{33}$ as the $Q_{33}$ is obtained by its derivative w.r.t. the compressive z-strain. Contrarily to the permitivity and $M_{33}$, a regular, non-divergent behaviour for both $\eta_{33}$ and $Q_{33}$.
In fact, when approaching the FEPT at -0.4\% epitaxial strain, the coefficient $Q_{33}$ evinces a constant behaviour, and whilst the coefficient decreases as the compressive strain increases beyond the transition, it is a very small change, i.e. $<3\%$ over the considered strain range. 

It might look surprising that the $M_{33}$ electrostrictive coefficient diverges at FEPT while the $Q_{33}$ is almost unaffected but this makes sense when we consider eqs.~\eqref{eq:Electro}, as $\eta=\nicefrac{1}{\chi}$ where we have that  $Q=-\nicefrac{1}{\chi^{2}}\nicefrac{\partial\chi}{\partial X}$.
Hence, the diverging susceptibility (i.e. large $\nicefrac{\partial\chi}{\partial X}$) is compensated by the $\nicefrac{1}{\chi^{2}}$ part such that they balance each other out to give a nearly constant $Q$ coefficient at the FEPT, while the $M$ coefficient is directly proportional to $\nicefrac{\partial\chi}{\partial X}$ and increases with it.
Therefore, whether or not electrostriction is claimed to diverge and become huge at the FEPT~\cite{IsSm89} or  not~\cite{UwSa75,LiJi14} depends on whether one analyses the $M$ and $m$ coefficients (which \emph{must} diverge), or the $Q$ and $q$ coefficients (which do not) for continuous transitions.

Finally, we demonstrate that the ferroelectric phase offers a unique approach to the calculation of the electrostrictive coefficients, not yet implemented in DFT. 
This is based on the relation of the electrostriction tensor to the spontaneous polarisation and piezoelectricity tensor:\cite{Devonshire}\footnote{See Supplemental Material at [URL will be inserted by publisher] for derivation.}
\begin{equation} \label{eq:NewQ}
    g_{ijm}=2Q_{ijmn}P_{n},
\end{equation} 
where $g_{ijm}$ is the piezoelectric tensor which describes the linear coupling of strains to polarisations.
This means that in the range of strain where KTaO$_3$ is ferroelectric (below -0.4\%), we can determine the $Q$ coefficient through the spontaneous polarization (calculated through the Berry phase method) and the piezoelectric coefficient (directly accessible from DFPT).
Just below the critical strain for FEPT (-0.4\%), we obtained a spontaneous polarisation of $P_{3}=-1.57$ $\mu C/cm^{2}$ and a piezoelectricity tensor components $g_{33}$ and $g_{31}$ of -4.015 and 1.035 10$^{-3}$m$^{2}$/C, respectively. 
Using eq.~\eqref{eq:NewQ}, this yields to $Q_{11}=0.1278$~m$^{4}$C$^{-2}$ and $Q_{12}=-0.03295$~m$^{4}$C$^{-2}$, in good agreement with the $Q$ values of 0.1278 and -0.02986~m$^{4}$C$^{-2}$, evaluated at the same point using eqs.\ref{eq:Electro}. 
This method to obtain the $Q_{ijkl}$ coefficients has advantages over both the methodology we have presented in Ref.\onlinecite{TaBo21}, and the finite field methodologies found in the literature~\cite{JiZh16,CaFo11,WaMe10}, the primary amongst which is that no finite differences or relaxations were required, just the ground-state based Berry phase and DFPT calculations. 
It also circumvents issues associated with applying an electric field to the paraelectric phase of a perovskite~\cite{JiZh16}, and more general shortcomings associated with finite field methodologies~\cite{TaBo21}.

In conclusion, we have performed a thorough investigation of electrostriction at the ferroelectric phase transition.
We have outlined a route towards giant electrostriction through epitaxial strain engineering
, demonstrating an electrostrictive $M_{33}$ coefficient equivalent to an effective $d_{33}$ coefficient of $60,000$~pm/V. 
The magnitude of this response, the ease with which the studied system may be grown, and the advantages of lead-free electrostrictive materials over piezoelectric materials more generally mark this as a significant scientific and technological advancement. 
In addition, through our treatment of both the $Q_{ijkl}$ and $M_{ijkl}$ tensors, we underline that 
whether one will observe a diverging and giant electrostriction, or a relatively constant and stable response, depends on whether one measures the $Q$ or the $M$ tensor. Finally, we have introduced a new method by which electrostriction may be computed in a ferroelectric using only the Berry phase technique and DFPT, without the need for relaxations or finite differences.
Our results also question if the origin of giant electrostriction~\cite{JiachPierre} e.g. in LAMOX, doped ceria, or Bi$_2$O$_3$, is coming from an intrinsic softening of a polar mode, like in KTaO$_3$ or if its due to intrinsic electronic effects or other extrinsic effects.
This will require further experimental and theoretical researches to further understand giant electrostrictors.

\section*{Acknowledgements}
Computational resources have been provided by the Consortium des \'Equipements de Calcul Intensif (C\'ECI), funded by the Fonds de la Recherche Scientifique (F.R.S.-FNRS) under Grant No. 2.5020.11 and by the Walloon Region and using the DECI resource BEM based in Poland at Wrocław with support from the PRACE OFFSPRING project.
EB acknowledges FNRS for support and DT aknowledge ULiege Euraxess support.
This work was also performed using HPC resources from the “M\'esocentre” computing centre of CentraleSup\'elec and \'Ecole Normale Sup\'erieure Paris-Saclay supported by CNRS and R\'egion \^Ile-de-France (http://mesocentre.centralesupelec.fr/).
Financial support is acknowledged from public grants overseen by the French National Research Agency (ANR) in the  ANR-20-CE08-0012-1 project and as part of the ASTRID program (ANR-19-AST-0024-02). 
This article is based upon work from COST Action OPERA - EurOPean NEtwork foR Innovative and Advanced Epitaxy, supported by COST (European Cooperation in Science and Technology.

\bibliographystyle{apsrev4-1}              
\bibliography     {./Bib} 

\end{document}


\title{}
\title{Supplementary Information: \\ 
Strain-engineered divergent electrostriction in \ce{KTaO3}}


\author{Daniel S.~P.~Tanner}
\email{danielsptanner@gmail.com} 
\affiliation{Universit\'e de Li\`ege, Q-MAT, CESAM, Institut de Physique}
\affiliation{Universit\'e Paris-Saclay, CentraleSup\'elec, CNRS, Laboratoire SPMS, 91190 Gif-sur-Yvette, France}
\author{Pierre-Eymeric Janolin}
\affiliation{Universit\'e Paris-Saclay, CentraleSup\'elec, CNRS, Laboratoire SPMS, 91190 Gif-sur-Yvette, France}
\author{Eric Bousquet}
\affiliation{Universit\'e de Li\`ege, Q-MAT, CESAM, Institut de Physique}
\email{email}

\maketitle

\section*{Thermodynamics of indirect electrostrictive effect}
\subsection*{Electrostriction as stress/strain derivative of permitivity/dielectric susceptibility}
To confirm the thermodynamic validity of the methodology we propose 
, we will derive it for the coefficients $Q_{ijkl}$. Start by considering the elastic Gibbs free energy for a bulk material:
\begin{eqnarray} \label{eq:elastic_Gibbs}
    G_1 =&& u -T.s-Xx \\
    dG_1 =&& -s.dT - x_{kl}.dX_{kl} + E_i.dP_i
\end{eqnarray}
where $u$ is the internal energy per unit volume (J/m$^3$), $T$ the temperature (K), $s$ the volume density of entropy (J/m$^3$), $E$ the electric field (V/m), and $P$ the polarisation (C/m$^2$ ). Taking first derivatives, and considering only electromechanical deformations, in linear dielectrics, we have:
\begin{eqnarray} 
    \frac{\partial G_1}{\partial X_{ij}} &= -x_{kl};~~    &x_{ij} =g_{ijk}P_{k} + Q_{ijkl}P_{k}P_{l}\label{eq:dGdX}\\ 
    \frac{\partial G_1}{\partial P_{i}}  &= -E_i;~~       &E_i    =\eta_{ij}P_{j} \label{eq:dGdD}
\end{eqnarray}
Where $\eta_{ij}$ is the dielectric stiffness. Now proceed to third derivatives by differentiating Eq.~(\ref{eq:dGdX}) by $P_{i}$ followed by $P_{j}$:
\begin{equation} \label{eq:D D X}
 \frac{\partial^{3}G_1}{\partial P_j\partial P_i \partial X_{kl}} =  -\frac{\partial^{2}x_{kl}}{\partial P_j\partial P_i} = -2Q_{ijkl}
\end{equation}
and differentiating Eq.~(\ref{eq:dGdD}) by $P_{j}$ and $X_{kl}$:
\begin{equation} \label{eq: X D D}
 \frac{\partial^{3}G_1}{\partial X_{kl}\partial P_j\partial P_i } = \frac{\partial^{3}G_1}{\partial X_{kl} \partial P_j\partial P_i } = \frac{\partial\eta_{ij}}{\partial X_{ij}}
\end{equation}
$G_1$ being $\mathcal{C^\infty}$, by Schwarz's theorem, we have that Eq.~(\ref{eq:D D X})=Eq.~(\ref{eq: X D D}) and the desired result, defining the converse electrostrictive effect: 
\begin{equation} \label{eq:converse_Q2}
  -\frac{1}{2}\frac{\partial\eta_{ij}}{\partial X_{ij}} = Q_{ijkl}  
\end{equation}

The expressions for the other electrostrictive coefficients, $q_{ijkl}$, $m_{ijkl}$, and $M_{ijkl}$ may be derived in a similar manner, but starting from the  Helmholtz free energy, the electric Gibbs free energy, and Gibbs free energy, respectively.

\subsection*{Electrostriction from piezoelectric tensor and spontaneous polarisation}
In a non-piezoelectric material at equilibrium, a strain will develop due to stresses, according to the elastic compliance, and due to polarisations, according to the electrostrictive Q tensor:
\begin{equation} \label{eq:strain}
  x_{i} = -s_{ij}X_{j} +Q_{ijmn} P_{m}P_{n} ,
\end{equation}
where Voigt notation has been used to simplify the tensors. Taking the derivative of this equation:
\begin{equation} \label{eq:strain2}
  \frac{\partial x_{i}}{\partial P_{m}} = 2Q_{ijmn}P_{n} .
\end{equation}
The left hand side of this equation defines the piezoelectric coefficient $g_{mi}$, meaning that a material which normally has
zero piezoelectric coefficients will have finite ones if a polarisation is induced, which, in the case of this paper, by straining to the ferroelectric phase.